\documentclass[prl,aps,twocolumn,showpacs]{revtex4}
\usepackage{graphicx}
\begin{document}
\title{Optical simulation of neutrino oscillations in binary waveguide arrays}
\author{Andrea Marini$^{1,*}$, Stefano Longhi$^2$ and Fabio Biancalana$^{1,3}$}
\email{Andrea.Marini@mpl.mpg.de, Fabio.Biancalana@mpl.mpg.de}
\affiliation{$^1$Max Planck Institute for the Science of Light, Guenther-Scharowsky-Stra\ss e 1, 91058 Erlangen, Germany}
\affiliation{$^2$Dipartimento di Fisica, Politecnico di Milano, and IFN-CNR, Piazza L. da Vinci 32, I-20133 Milano, Italy}
\affiliation{$^3$School of Engineering \& Physical Sciences, Heriot-Watt University, Edinburgh, EH144AS, United Kingdom}
\date{\today}
\begin{abstract}
We theoretically propose and investigate an optical analogue of neutrino oscillations in a pair of
vertically displaced binary waveguide arrays with longitudinally modulated effective refractive index.
Optical propagation is modelled through coupled-mode equations, which in the continuous limit 
lead to two coupled Dirac equations for fermionic particles with different mass states, i.e. neutrinos. 
We demonstrate that neutrino oscillations can be quenched by nonlinear effects, and we predict the existence of 
neutrino solitons. Incidentally, these phenomena are expected to
play an important role in massive supernova stars. Our results pave the way for using binary waveguide arrays 
as a classical laboratory for predicting exotic effects in particle physics and astrophysics.

\end{abstract}
\maketitle

\paragraph{Introduction --} 

Optical waveguide arrays (WAs) allow for the optical simulation of non-relativistic dynamics of quantum particles
\cite{ChristoNat2003,LonghiLasPhotRev2009,GaranovichPhysRep2012}. They constitute a useful classical 
laboratory for mimicking quantum effects and can be exploited to analyse fundamental quantum mechanisms 
with classical tools, such as Bloch oscillations \cite{PertschPRL1999}, Zener tunnelling \cite{GhulinyanPRL2005,TrompeterPRL2006}, 
dynamical localisation \cite{LonghiPRL2006}, and Anderson localisation in disordered lattices \cite{LahiniPRL2008}. 
In addition, relativistic phenomena of quantum field theory can be optically simulated in binary waveguide arrays 
(BWAs), as optical propagation in the continuous limit is governed by a (1+1)D Dirac equation \cite{LonghiPRB2010,DreisowPRL2010,LonghiAPB2011,ZeunerPRL2012}. 
Thus, BWAs can be engineered in order to observe diverse relativistic effects, e.g. 
Zitterbewegung \cite{DreisowPRL2010}, Klein tunnelling \cite{LonghiPRB2010,DreisowEPL2012}, fermion pair production \cite{DreisowPRL2012}, 
and solitons \cite{TranPRL2013}. Spontaneous symmetry breaking induced by tachyon condensation can be simulated in 
amplifying plasmonic arrays \cite{MariniPRA2014}, where optical propagation in the continuous limit satisfies a 
(1+1)D Dirac-like equation for particles with imaginary mass, i.e. tachyons. WAs aimed to simulate other 
quantum field theoretical models and non-physical particles, such as Majorana physics, have been suggested as 
well \cite{Maj1,Maj2}. Besides optical WAs, 
quantum simulations in trapped ion systems have provided spectacular results in this context \cite{trapped1,trapped2,trapped3,trapped4}.\\
An important mechanism arising 
in particle physics is represented by {\it neutrino oscillation} \cite{Pontecorvo1957}, a quantum mechanical effect 
whereby a neutrino created with a specific lepton flavor can later be measured with a different flavor. Neutrino 
oscillations occur both in vacuum and in matter, as the probability of measuring a neutrino with a specific flavor 
periodically oscillates during the propagation \cite{Wolfenstein1978,Kuo1989,Gonzalez2003}. 
In recent years, theoretical and experimental investigations of neutrino oscillations have received considerable 
interest, as the observation of this phenomenon implies that neutrinos have small but finite masses. While in the 
Standard Model of particle physics the mass of charged fermions results from the interactions with the Higgs field, 
the origin of the neutrino masses has not hitherto been answered conclusively. Neutrino oscillations ensue from a 
mixture between the flavor and mass eigenstates, since the three neutrino states weakly interacting with charged leptons 
are superpositions of three states with definite mass. \\
The enormous neutrino fluxes emitted by supernovae are key to the explosion dynamics and nucleosynthesis \cite{Janka2007} 
and detecting a high statistics of neutrinos from the next nearby supernova explosion is a major aim of neutrino astronomy. 
In the early universe and in collapsing supernova stars, the density of neutrinos is so large that they strongly interact 
with each other, many body interactions become important and neutrino oscillations feedback on themselves 
\cite{Pantaleone1992,Vissani2010,Duan2010,Balantekin2013}. In turn, peculiar effects like coherence in collective neutrino oscillations 
and suppression of self-induced flavor conversion occur in supernova explosions \cite{Raffelt2011,Sarikas2012}.
Owing to the experimental difficulty of investigating neutrino oscillations in these extreme conditions, quantum 
simulation with trapped ions has been recently proposed \cite{NohNJP2012}.

In this Letter we theoretically investigate optical propagation in a pair of vertically displaced binary waveguide 
arrays with longitudinally modulated effective index. In the fast modulation and continuous limit, light dynamics 
is described by two coupled nonlinear Dirac equations for fermionic particles with different mass states, analogously to
the equations governing the evolution of neutrinos in matter \cite{Gonzalez2003}. In the linear limit where nonlinear effects 
are negligible, our optical setup allows for the simulation of neutrino oscillations, thoroughly reproducing the 
Pontecorvo-Maki-Nakagawa-Sakata (PMNS) Matrix formulation and the matter mixing matrix \cite{Maki1962,Pontecorvo1968}. 
In the nonlinear regime, physically accessible at high matter densities, e.g. in the early universe and in supernova stars, 
we predict the quenching of neutrino oscillations and the existence of topological defects, i.e. neutrino solitons.

\paragraph{Model --} 

In the following we consider a pair of vertically displaced BWAs with longitudinally modulated effective refractive index, 
sketched in Fig. \ref{Fig1}. The index modulation is assumed small and sinusoidal, and can be achieved by 
longitudinally modulating the core index or the transverse waveguide section (see, for instance, \cite{SzameitPRL2009}). 
We indicate by $A_n$, $B_n$ the field amplitudes of the linear fundamental modes of the upper and lower waveguides 
(see Fig. \ref{Fig1}a), which are weakly affected by the longitudinal modulation of the effective index. Typically, 
we assume that BWAs are manufactured in silica glass. Taking into account Kerr nonlinearity, nearest-neighbour evanescent 
coupling and longitudinal modulation of the effective index, optical propagation is described by the set of coupled-mode 
equations (CMEs): 

\begin{figure}[t]
\centering
\begin{center}
\includegraphics[width=0.235\textwidth]{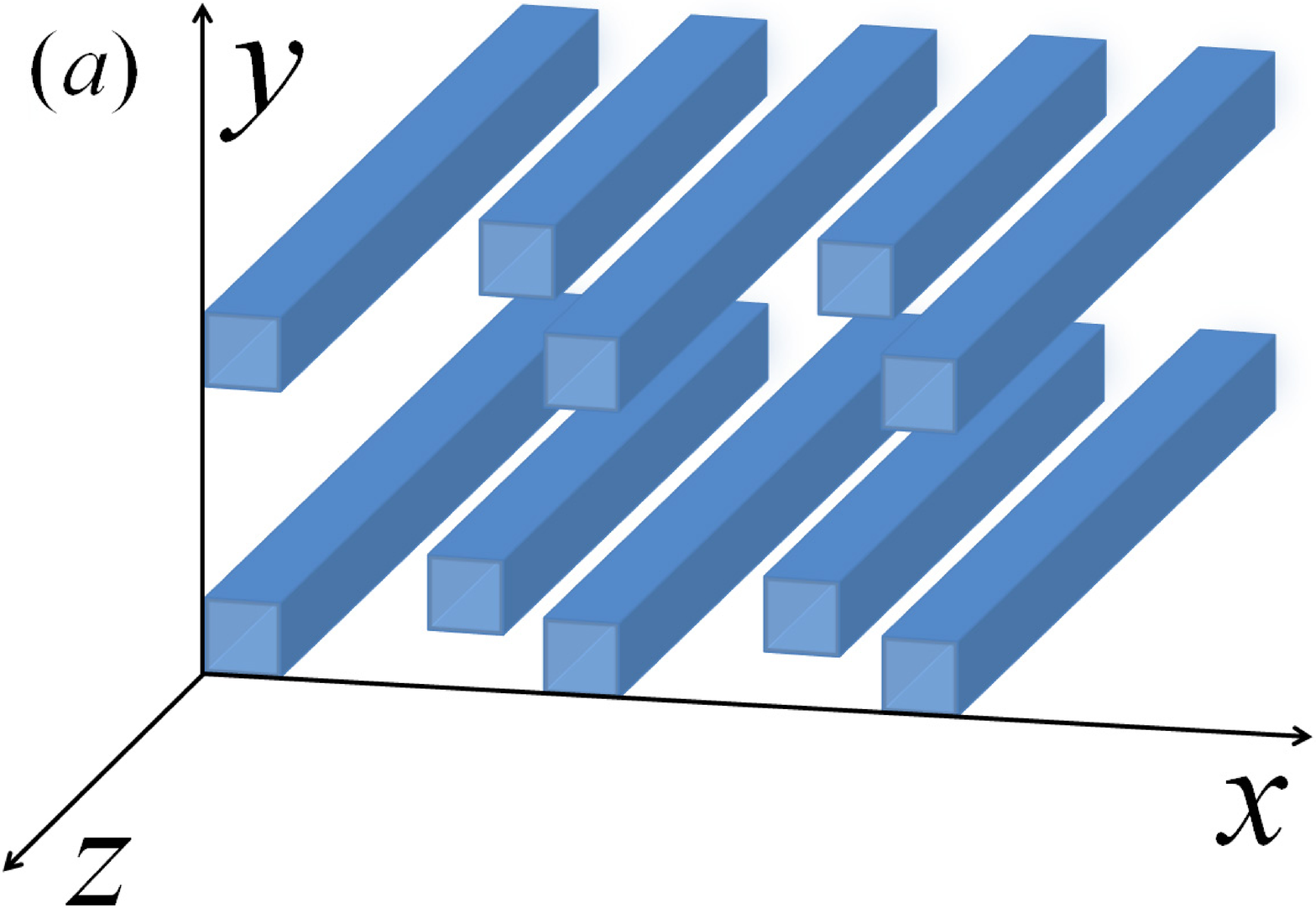}
\includegraphics[width=0.235\textwidth]{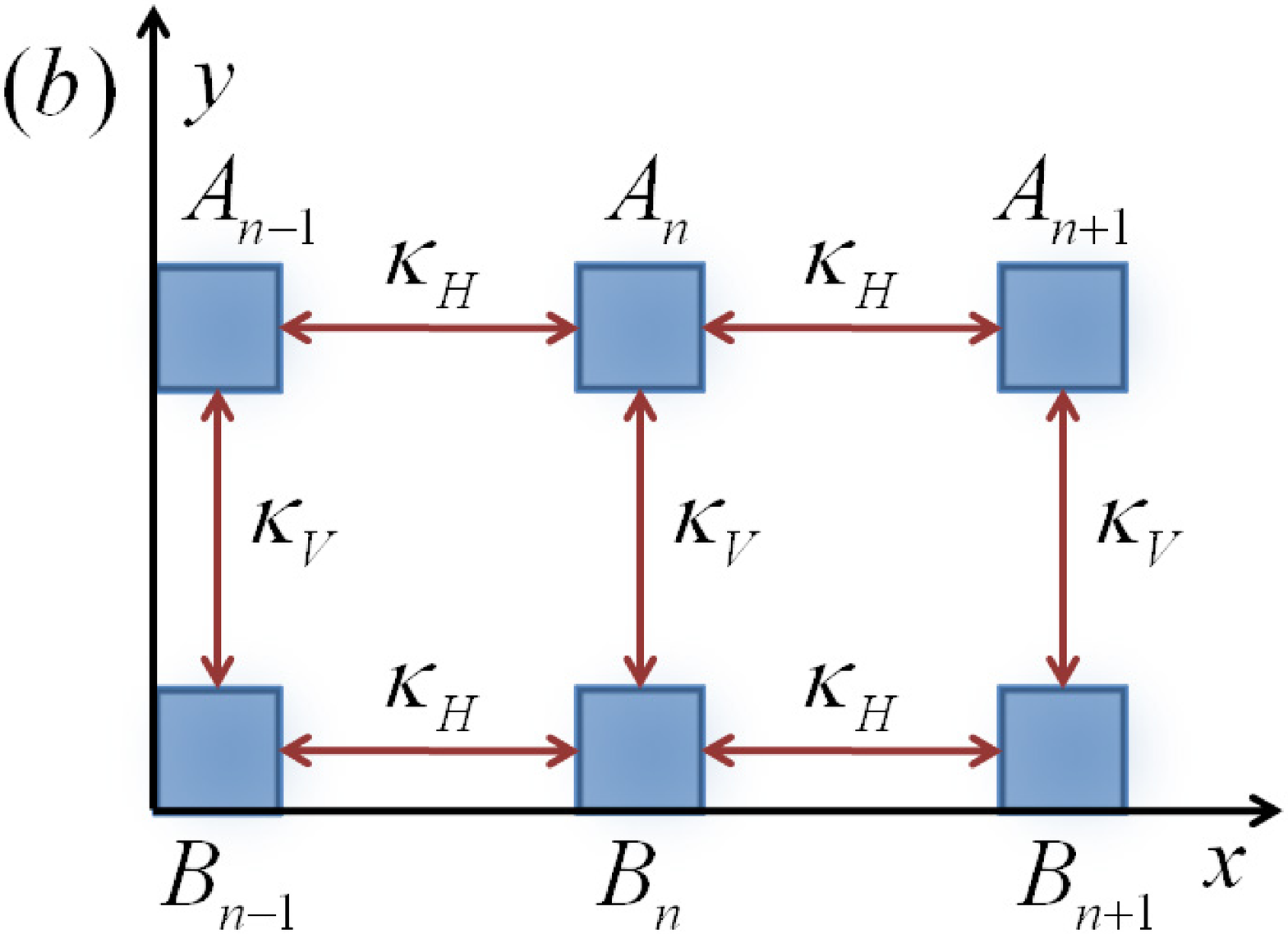}
\caption{Schematics of two vertically-displaced  BWAs for the simulation of neutrino oscillation. (a) Three-dimensional sketch of the structure. Neighbouring silica waveguides are offset, while every waveguide shows a longitudinally-modulated effective index. Light is 
trapped by every waveguide in the transverse $x$-$y$ directions and propagates in the longitudinal $z$-direction. 
(b) Transverse section of the structure. Optical amplitudes of the upper ($A_n$) and lower ($B_n$) array are 
coupled to nearest neighbours. Horizontal and vertical coupling constants between nearest neighbour waveguides 
are denoted by $\kappa_H,\kappa_V$, respectively.}
\label{Fig1}
\end{center}
\end{figure}

\begin{eqnarray}
&& i \frac{ d A_n}{ d z} = V_n(z) A_n - \kappa_H ( A_{n+1} + A_{n-1} ) - \kappa_V B_n + \nonumber \\
&& - \gamma | A_n |^2 A_n , \label{Eq1}\\
&& i \frac{ d B_n}{ d z} = W_n(z) B_n - \kappa_H ( B_{n+1} + B_{n-1} ) - \kappa_V A_n + \nonumber \\
&& - \gamma | B_n |^2 B_n, \label{Eq2}
\end{eqnarray}
where $n$ is an integer labelling the waveguide site (see Fig. \ref{Fig1}b), $\kappa_H,\kappa_V$ are the horizontal 
and vertical coupling coefficients between neighbouring waveguides (see Fig. \ref{Fig1}b), $\gamma$ is the Kerr 
nonlinear coefficient, $V_n(z) = (-1)^n m_a + \Delta V_n(z)$, $W_n(z) = - (-1)^n m_b + \Delta W_n(z)$, $m_a$ ($m_b$) 
account for the alternating initial offsets of the upper (lower) BWAs, and $\Delta V_n(z),\Delta W_n(z)$ describe the 
longitudinal modulation of propagation constants of the upper and lower waveguides. 
For a sinusoidal index modulation, we may write $\Delta V_n(z) = A_1 \mathrm{cos}(\omega z)$, 
$\Delta W_n(z) = A_3 \mathrm{cos}(\omega z)$ ($n$ even), $\Delta V_n(z) = A_2 \mathrm{cos}(\omega z)$, 
$\Delta W_n(z) = A_4 \mathrm{cos}(\omega z)$ ($n$ odd). After setting 
$A_n(z) = a_n(z)\mathrm{exp}\left\{ - i \int_0^z \Delta V_n(z') dz' \right\}$,
$B_n(z) = b_n(z)\mathrm{exp}\left\{ - i \int_0^z \Delta W_n(z') dz' \right\}$, in the fast modulation limit
$\omega \gg \kappa_H, \kappa_V , |m_{a,b}|$, at leading order we can average Eqs. (\ref{Eq1},\ref{Eq2}) and disregard 
the rapidly oscillating terms \cite{LonghiPRB2008}. This yields
\begin{eqnarray}
&& i \frac{ d a_n}{ d z} = - \kappa ( a_{n+1} + a_{n-1} ) + (-1)^n m_a a_n + \nonumber \\
&& + (-1)^n \epsilon b_n - \gamma | a_n |^2 a_n , \label{Eq3} \\
&& i \frac{ d b_n}{ d z} = - \kappa ( b_{n+1} + b_{n-1} ) - (-1)^n m_b b_n + \nonumber \\
&& + (-1)^n \epsilon a_n - \gamma | b_n |^2 b_n , \label{Eq4}
\end{eqnarray}
where $\kappa = \kappa_H J_0 [(A_2-A_1)/\omega] = \kappa_H J_0 [(A_4-A_3)/\omega]$,  
$\epsilon = \kappa_V J_0 [(A_3-A_1)/\omega] = - \kappa_V J_0 [(A_4-A_2)/\omega]$
are the effective averaged coupling constants between adjacent waveguides in the horizontal and vertical directions, 
respectively, and $J_0(x)$ is the zeroth-order Bessel function of the first kind. 
\begin{figure}[t]
\centering
\begin{center}
\includegraphics[width=0.235\textwidth]{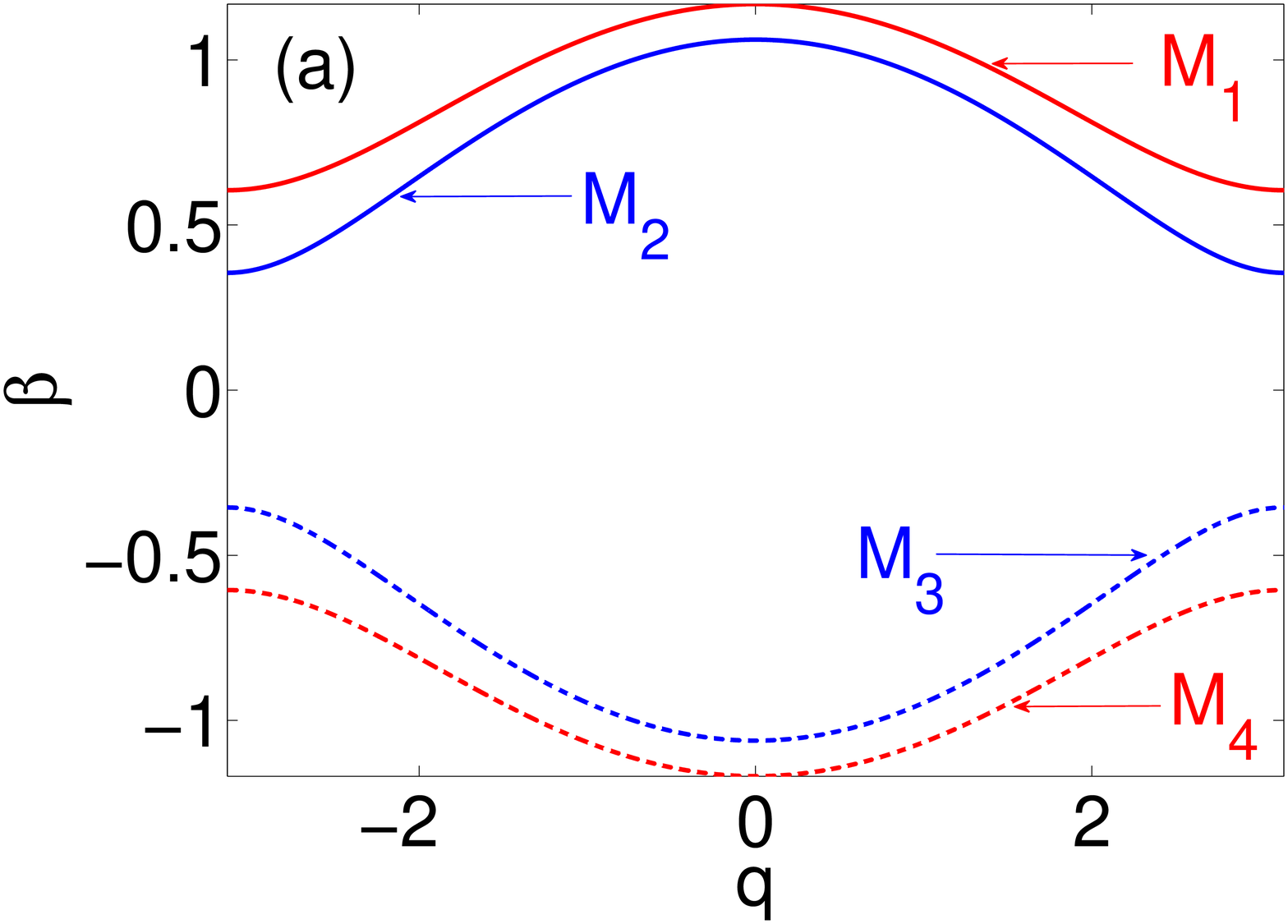}
\includegraphics[width=0.235\textwidth]{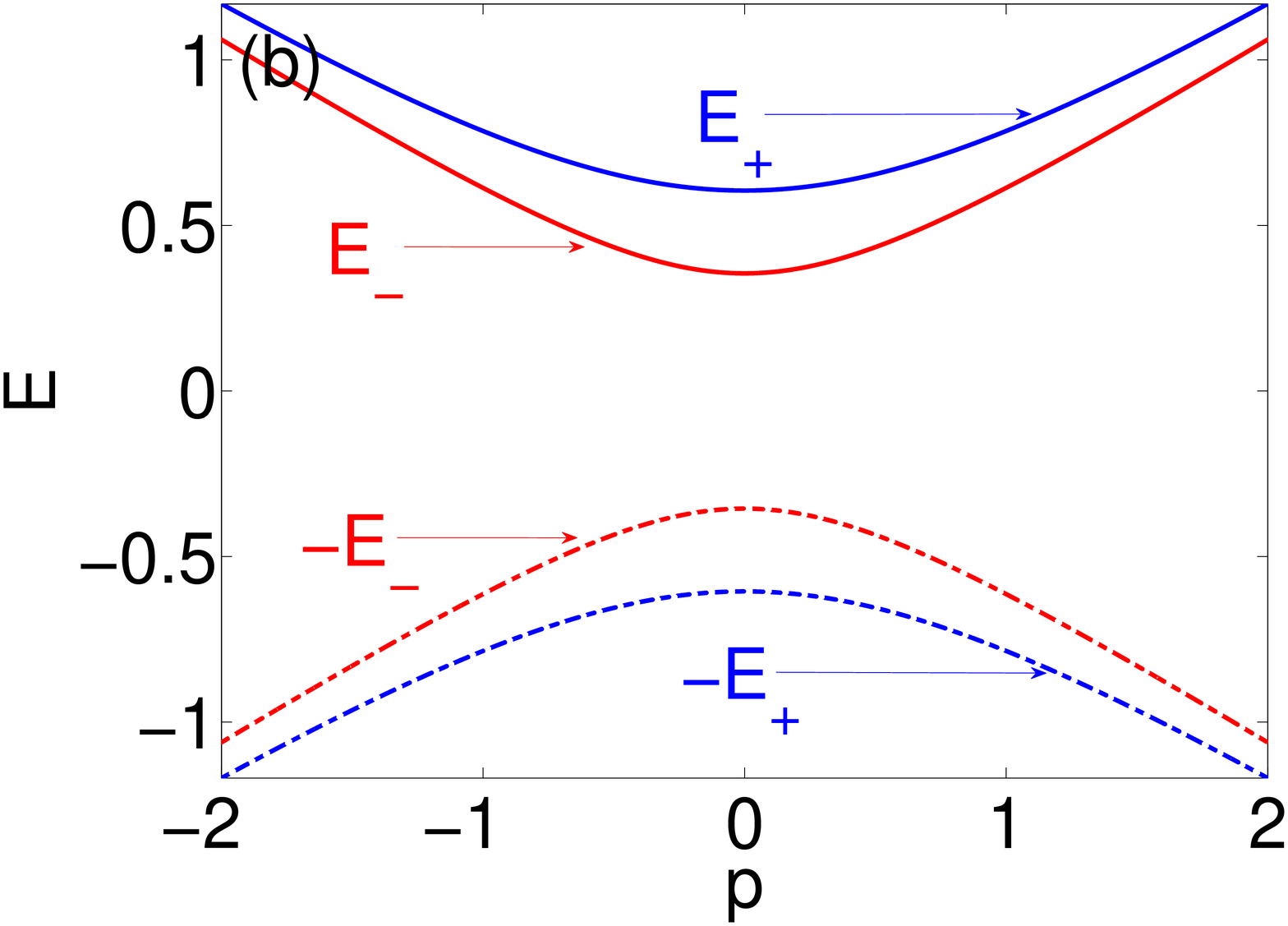}
\caption{(a) Linear dispersion of the structure sketched in Fig. \ref{Fig1}, supporting the linear modes 
$M_1,M_2,M_3,M_4$. (b) Linear dispersion of instantaneous mass eigenstates of neutrinos: energy $E$ as a function of momentum $p$. 
Full (dashed) lines represent the dispersion of neutrino (antineutrino) eigenstates, while blue (red) lines 
represent the dispersion of eigenstates with energies $E_+$ ($E_-$). Parameter values are $\kappa = 0.5$, $\epsilon = 0.3$, 
$m_a = 0.5$, $m_b = 0.25$.}
\label{Fig2}
\end{center}
\end{figure}
To obtain Eqs. (\ref{Eq3},\ref{Eq4}), we have assumed that the modulation amplitudes 
$A_1,A_2,A_3,A_4$ satisfy the constraints $J_0 [(A_2-A_1)/\omega] = J_0 [(A_4-A_3)/\omega]$, 
$J_0 [(A_3-A_1)/\omega] = - J_0 [(A_4-A_2)/\omega]$, which are readily fulfilled by taking $A_1 = 0$, 
$A_2 = (\omega/2)(\xi_1-\xi_2)$, $A_3 = \omega\xi_1$, $A_4 = (\omega/2)(\xi_1+\xi_2)$, where $\xi_1$, 
$\xi_2$ are two arbitrary dummy variables with $J_0(\xi_1)=-J_0(\xi_2)$. For instance,  one can take $\xi_1 = 1.6965$, $\xi_2 = 3.7152$ 
at which $J_0(\xi_1) = - J_0(\xi_2) = 0.4$, so that one obtains $\kappa = \kappa_H J_0(\xi_1/2-\xi_2/2) \simeq 0.761 \kappa_H$, 
$\epsilon = \kappa_V J_0(\xi_1) \simeq 0.4 \kappa_V$. Linear modes of Eqs. (\ref{Eq3},\ref{Eq4}) can be 
calculated by setting $\gamma = 0$ and by taking the Ansatz $a_{2n} = a_1 e^{i\beta z + i q n}$, 
$a_{2n+1} = a_2 e^{i\beta z + i q n}$, $b_{2n} = b_1 e^{i\beta z + i q n}$, $b_{2n+1} = b_2 e^{i\beta z + i q n}$, 
where $q$ is the transverse quasi-momentum in the $x$-direction. After substitution of
the Ansatz above in Eqs. (\ref{Eq3},\ref{Eq4}), one obtains the modes and the linear dispersion relation of the 
structure, which is depicted in Fig. \ref{Fig2}a. Note that our structure, consisting of a pair of vertically 
displaced BWAs, supports four linear modes $M_1,M_2,M_3,M_4$, characterized by four dispersion bands 
(see Fig. \ref{Fig2}a).

\begin{figure}[t]
\centering
\begin{center}
\includegraphics[width=0.235\textwidth]{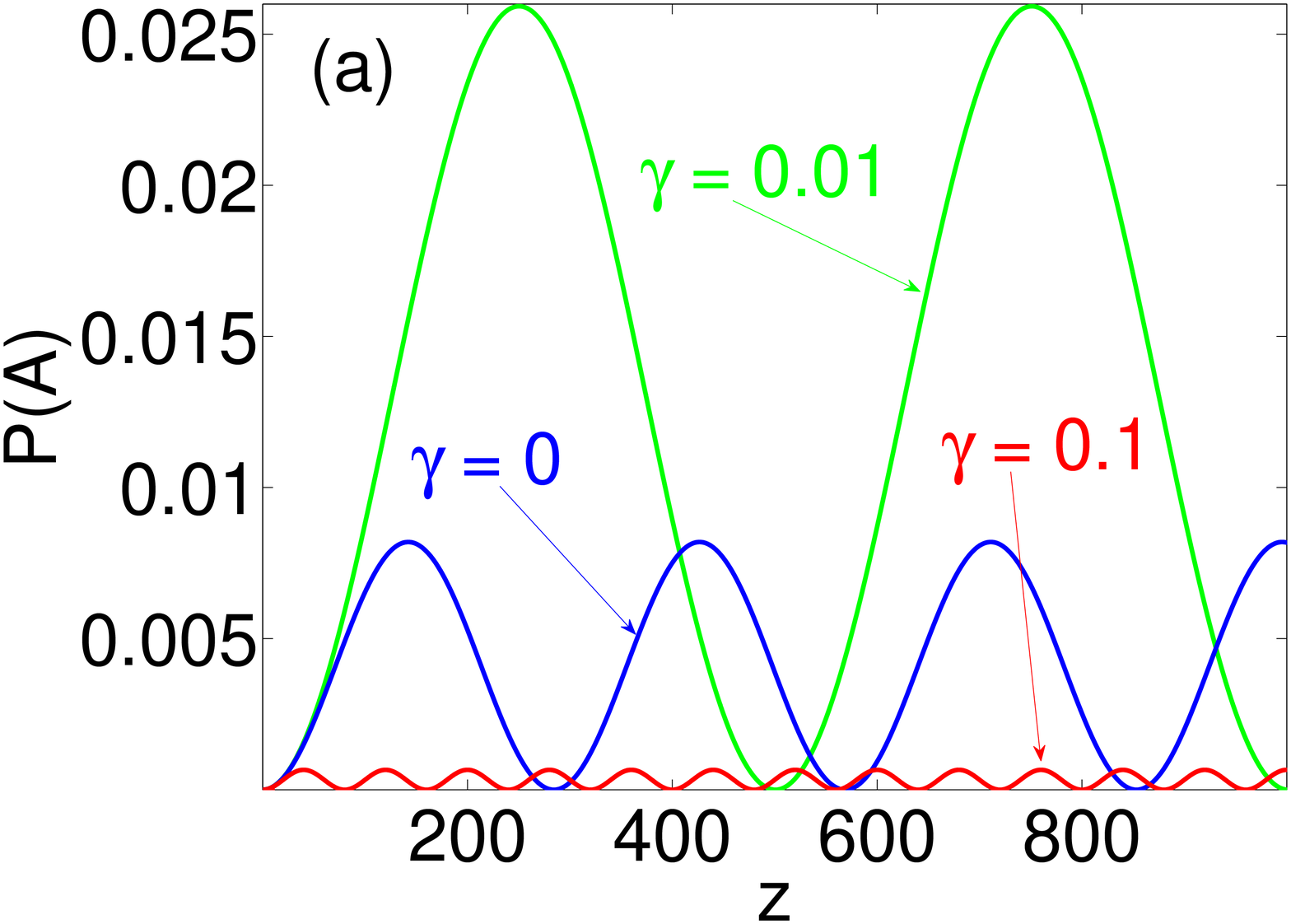}
\includegraphics[width=0.235\textwidth]{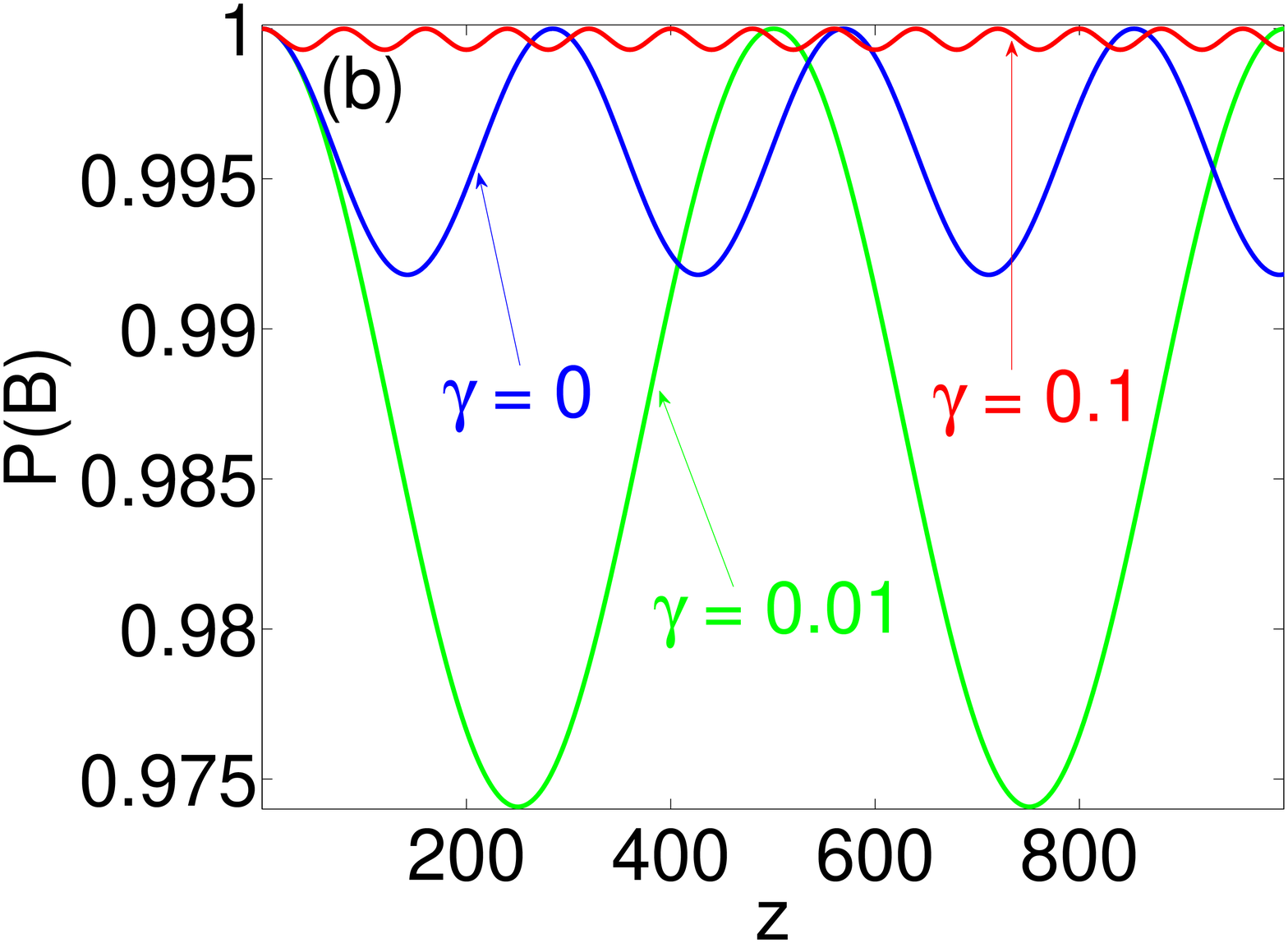}
\caption{Probabilities $P(A)$ [panel(a)] and $P(B)$ [panel (b)] of measuring the neutrino with flavors $A$
and $B$, respectively, as functions of the propagation length for several nonlinear
coefficients $\gamma = 0, 0.01,0.1$, corresponding to the blue, green and red curves. The other parameter values 
are $\kappa = 1$, $\epsilon = 0.001$, $m_a = 0.01$, $m_b = 0.012$.}
\label{Fig3}
\end{center}
\end{figure}

\paragraph{Dirac limit: neutrinos --}

Defining the two-component spinors 
$A = (A_1 ~ A_2)^T = (-1)^n ( a_{2n} , i a_{2n-1} )^T$ and $B = (B_1 , B_2)^T = (-1)^n ( i b_{2n-1} , b_{2n} )^T$,
if the amplitudes $A_1$, $A_2$, $B_1$, $B_2$ vary slowly with the site index $n$, one can take the continuous limit 
by introducing the continuous spatial coordinate $n \rightarrow x$ \cite{LonghiPRB2010}. In this limit, the spinors satisfy two coupled 
(1+1)D nonlinear Dirac equations for half-spin particles with two different mass eigenstates, i.e. neutrinos:
\begin{eqnarray}
&& i \partial_z A = - i \kappa \hat{\sigma}_x \partial_x A + m_a \hat{\sigma}_z A + i \epsilon \hat{\sigma}_y B - \gamma G ( A ) , \;\;\; \label{NLDiracEq1} \\
&& i \partial_z B = - i \kappa \hat{\sigma}_x \partial_x B + m_b \hat{\sigma}_z B - i \epsilon \hat{\sigma}_y A - \gamma G ( B ) , \;\;\; \label{NLDiracEq2}
\end{eqnarray}
where $G ( \psi ) \equiv ( |\psi_1|^2 \psi_1  , |\psi_2|^2 \psi_2 )^T$ is the nonlinear spinorial term and 
$\hat{\sigma}_x$, $\hat{\sigma}_y$, $\hat{\sigma}_z$ are the $x$, $y$, $z$ Pauli matrices. In the Dirac limit, the 
array alternating offsets $m_a,m_b$ play the role of the neutrino masses, while the coupling coefficients 
$\kappa,\epsilon$ play the role of the speed of light in vacuum and of the charged-current electroweak interactions.  
Remarkably, the linear parts of Eqs. (\ref{NLDiracEq1},\ref{NLDiracEq2}) are identical to models routinely
used in particle physics for describing neutrino oscillations in matter \cite{Gonzalez2003}. Nonlinear terms
are usually disregarded as neutrinos interact weakly in standard conditions of matter densities. However, recent studies
demonstrate that nonlinearity plays an important role in supernova stars and in the early universe, where matter
density is enormous \cite{Pantaleone1992,Balantekin2013,Raffelt2011,Sarikas2012}. The linear supermodes 
$|\psi_+\rangle$, $|\psi_-\rangle$ of Eqs. (\ref{NLDiracEq1},\ref{NLDiracEq2}) (calculated by setting $\gamma =0$) 
represent the instantaneous mass neutrino eigenstates in matter \cite{Gonzalez2003}, which can be expressed as 
linear superpositions of the mass eigenstates $|M_a\rangle$, 
$|M_b\rangle$, i.e. the linear modes of Eqs. (\ref{NLDiracEq1},\ref{NLDiracEq2}) in the uncoupled limit 
$\epsilon\rightarrow 0$ (limit of non-interacting neutrinos, e.g. in vacuum):
\begin{eqnarray}
&& |\psi_+ \rangle = \mathrm{sin} \Theta |M_a\rangle - \mathrm{cos} \Theta |M_b\rangle, \label{Eq7} \\
&& |\psi_- \rangle = \mathrm{cos} \Theta |M_a\rangle + \mathrm{sin} \Theta |M_b\rangle, \label{Eq8}
\end{eqnarray}
where $\Theta = \mathrm{tan}^{-1}[2\epsilon/(m_a+m_b)]/2$ is the mixing angle. Note that Eqs. (\ref{Eq7},\ref{Eq8}) 
coincide with the neutrino mixing matrix in matter \cite{Gonzalez2003}. The supermodes $|\psi_+\rangle$, 
$|\psi_-\rangle$ can be calculated with the Ansatz $A = {\cal A} e^{i E z + i p x}$, 
$B = {\cal B} e^{i E z + i p x}$, which yields the dispersion relation 
$E_{\pm}^2 = \kappa^2 p^2 + \epsilon^2 + (m_a^2 + m_b^2)/2 \pm (m_b - m_a)\sqrt{\epsilon^2+(m_a+m_b)^2/4}$,
plotted in Fig. \ref{Fig2}b. Curves with positive (negative) energies represent the instantaneous mass 
dispersion of neutrinos (antineutrinos). Note that the four dispersion curves of Fig. \ref{Fig3} basically 
reproduce the BWAs dispersion relations of Fig. \ref{Fig2} near $q=\pi$.  

\begin{figure}[t]
\centering
\begin{center}
\includegraphics[width=0.5\textwidth]{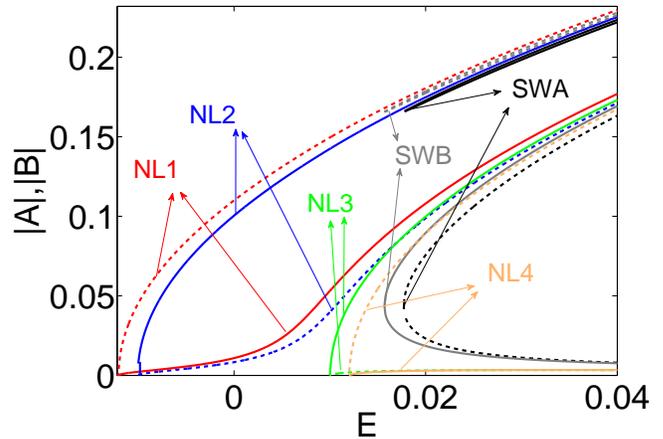}
\caption{Homogeneous nonlinear mode families not undergoing oscillation. Full (dashed) curves show
the modulus of the spinor amplitudes $|A|$ ($|B|$) as functions of the energy $E$ (propagation constant in our 
optical analogy). Parameter values are $\kappa = 1$, $\epsilon = 0.001$, $m_a = 0.01$, 
$m_b = 0.012$, $\gamma = 1$.}
\label{Fig4}
\end{center}
\end{figure}

\paragraph{Neutrino oscillations} Eqs. (\ref{NLDiracEq1},\ref{NLDiracEq2}), analogously to models routinely used
in particle physics \cite{Gonzalez2003}, describe neutrino oscillations in matter. Neutrinos are created in weak 
processes in their flavor eigenstates. As they propagate, the quantum mechanical phases of the mass states flow
at different rates owing to the diverse neutrino masses. Analogously to Eqs. (\ref{NLDiracEq1},\ref{NLDiracEq2}), 
also neutrino flavors can be expressed in terms of mass states through the PMNS matrix formulation of flavor mixing 
\cite{Maki1962,Pontecorvo1968}. In turn, neutrino oscillations in vacuum are trivial, in the sense that they are 
simply given by the beating between mass states with different energies (propagation constants in our optical 
analogy). In matter, neutrino oscillations are due to electroweak interactions, which in our optical setup are 
accounted for by the coupling coefficient $\epsilon$. For instance, considering a neutrino that at $z=0$ is in 
the electron flavor state, and initially neglecting nonlinear effects, the probability of measuring the neutrino 
with electron and muon flavors oscillates in $z$: 
$P(A) = \mathrm{sin}^2(2\Theta)\mathrm{sin}^2[(E_+ - E_-)z/2]$,
$P(B) = 1 - \mathrm{sin}^2(2\Theta)\mathrm{sin}^2[(E_+ - E_-)z/2]$. In principle, the oscillation mixing angle 
results from the composition of the PMNS matrix angle and electroweak interactions, giving rise to peculiar phenomena
such as the Mikheyev-Smirnov-Wolfenstein resonance \cite{Wolfenstein1978}. However, in our calculations we have 
considered only the effect of electroweak interactions, accounted for by the mixing angle $\Theta$, focusing directly
on the oscillations of the mass states. In massive supernova stars, matter density is
so high that electroweak interactions become nonlinear, affecting neutrino oscillations. We have numerically solved 
Eqs. (\ref{NLDiracEq1},\ref{NLDiracEq2}) for homogeneous waves with null transverse momentum using a fourth order 
Runge Kutta algorithm. Results of numerical simulations for several values of the nonlinear coefficient $\gamma$ 
are shown in Figs. \ref{Fig3}(a,b). The probabilities $P(A),P(B)$ of measuring neutrinos with flavors $A,B$ 
oscillate in $z$ both in the linear and nonlinear regimes. Note that the effect of nonlinearity on the amplitude 
and the period of the oscillations is nontrivial. Indeed, as the nonlinear coefficient $\gamma$ increases, the 
oscillation amplitude and period initially increase for $\gamma = 0.01$ (see the green and blue curves of 
Figs. \ref{Fig3}a,b). Conversely, oscillations are quenched for the higher nonlinear coefficient $\gamma = 0.1$ 
[see the red and green curves of Figs. \ref{Fig3}(a,b)]. In order to grasp a better understanding of the nonlinear 
quenching of neutrino oscillation, we have calculated the homogeneous nonlinear mode families of 
Eqs. (\ref{NLDiracEq1},\ref{NLDiracEq2}) (nonlinear neutrino states not undergoing oscillation) by using the 
Newton-Raphson method. We focused our attention on nonlinear modes with null transverse momentum, and numerically 
found several neutrino families, which are depicted in Fig. \ref{Fig4}. The families $NL1,NL2$ bifurcate from the 
linear antineutrino bands with negative energy, while other families represent neutrinos with positive energies. 
The nonlinear families $NL3,NL4$ exist for any neutrino amplitude as they bifurcate from the linear neutrino bands, 
while the switching families $SWA,SWB$ ensue after a certain power threshold. In turn, the nonlinear quenching of 
neutrino oscillations comes from the excitation of the switching families $SWA,SWB$, which arise when nonlinearity 
is sufficiently large.

\paragraph{Neutrino solitons --}

\begin{figure}[t]
\centering
\begin{center}
\includegraphics[width=0.5\textwidth]{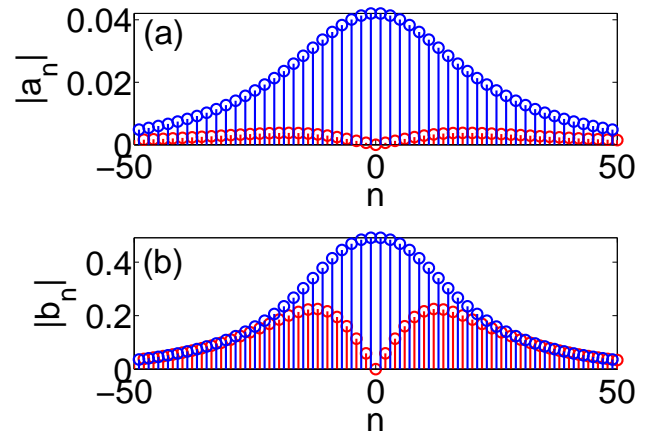}
\caption{Coupled neutrino-antineutrino soliton with null energy $E=0$ (propagation constant in our optical analogy). 
Blue (red) markers depict the amplitude of odd (even) waveguide sites, related to the spinor component $A_2$ ($A_1$)
in the continuous limit. In our calculations we have used the parameters $\kappa = 1$, $\epsilon = 0.01$, $m_a = 0.1$, 
$m_b = 0.12$, $\gamma = 1$.}
\label{Fig5}
\end{center}
\end{figure}

Owing to nonlinear electroweak interactions, another class of neutrino fields homotopically distinct from homogeneous 
ones exists, i.e. solitons arising from the condensation of neutrinos and antineutrinos. We have numerically solved 
Eqs. (\ref{Eq3},\ref{Eq4}) by taking the Ansatz $a_n(z) = {\cal A}(n) e^{i E z}$, 
$b_n(z) = {\cal B}(n) e^{i E z}$ and using the Newton-Raphson algorithm to find the soliton spinor profiles 
${\cal A}(n),{\cal B}(n)$. We have found several families of neutrino solitons within the energy 
bandgap (see Fig. \ref{Fig2}b). High-order solitons are cut-off when the energy bandgap is sufficiently small. 
In Fig. \ref{Fig5} we plot the fundamental soliton profiles with null energy $E=0$ of the (a) upper ($|a_n|$) and (b) 
lower ($|b_n|$) BWAs. Blue (red) markers depict the amplitude of odd (even) waveguide sites, related to the spinor 
component $A_2$ ($A_1$) in the continuous limit. In our calculations we have used the parameters $\kappa = 1$, 
$\epsilon = 0.01$, $m_a = 0.1$, $m_b = 0.12$, $\gamma = 1$. We studied the stability of this soliton-like solution in 
propagation under small perturbations, and found that it is stable. A more detailed analysis on the modulation 
instability of homogeneous modes and on the existence and stability conditions of solitons goes beyond the scope of 
this Letter and it will be presented elsewhere.

\paragraph{Conclusions --}

In conclusion we have shown that the rich physics of neutrino oscillations, including extreme regimes predicted in 
massive supernova stars, can be simulated in a tabletop optical system based on pairs of vertically displaced 
binary waveguide arrays. In this system light transport  is described by nonlinear (1+1)D Dirac equations for half-spin 
particles with two mass states, analogously to neutrinos. In the linear regime, our optical system exactly reproduces 
all the features of neutrino oscillations, including the Pontecorvo-Maki-Nakagawa-Sakata matrix formulation and the matter
mixing matrix. We predict 
the quenching of neutrino oscillations when electroweak interactions are sufficiently strong to become nonlinear, e.g. in 
the early universe and in massive supernova stars. Besides, we predict the condensation of neutrinos and antineutrinos  
in soliton pairs with energy within the linear bandgap. Our results show that optical waveguide arrays can provide an 
experimentally accessible laboratory tool for the observation of unconventional effects and extreme regimes in particle 
physics and astrophysics that are still out of any experimental demonstration.

This research is supported by the German Max Planck Society for the Advancement of Science (MPG).
A. M. acknowledges fruitful discussions with Luca Di Luzio.

\end{document}